\documentclass[pdflatex,sn-mathphys-ay]{sn-jnl}

\usepackage{etoolbox} 

\makeatletter
\patchcmd{\NAT@citex}
  {\@citea\NAT@hyper@{%
     \NAT@nmfmt{\NAT@nm}%
     \hyper@natlinkbreak{\NAT@aysep\NAT@spacechar}{\@citeb\@extra@b@citeb}%
     \NAT@date}}
  {\@citea\NAT@nmfmt{\NAT@nm}%
   \NAT@aysep\NAT@spacechar\NAT@hyper@{\NAT@date}}
  {}
  {}
\makeatother

\usepackage{graphicx}%
\usepackage{multirow}%
\usepackage{amsmath,amssymb,amsfonts}%
\usepackage{amsthm}%
\usepackage{mathrsfs}%
\usepackage[title]{appendix}%
\usepackage{xcolor}%
\usepackage{textcomp}%
\usepackage{manyfoot}%
\usepackage{booktabs}%
\usepackage{algorithm}%
\usepackage{algorithmicx}%
\usepackage{algpseudocode}%
\usepackage{listings}%
\usepackage{amsthm}%
\usepackage{centernot}%
\usepackage{float}%

\theoremstyle{thmstyleone}%
\newtheorem{theorem}{Theorem}
\theoremstyle{thmstyletwo}%
\newtheorem{lemma}[theorem]{Lemma}%

\theoremstyle{thmstylethree}%

\begin{document}

\title[Article Title]{Source Detection in Hypergraph Epidemic Dynamics using a Higher-Order Dynamic Message Passing Algorithm}


\author[1]{\fnm{Qiao} \sur{Ke}}
\author[3,4,5]{\fnm{Naoki} \sur{Masuda}}
\author[6]{\fnm{Zhen} \sur{Jin}}
\author[1]{\fnm{Chuang} \sur{Liu}}
\author*[1,2]{\fnm{Xiu-Xiu} \sur{Zhan}}\email{zhanxiuxiu@hznu.edu.cn}

\affil[1]{\orgdiv{Alibaba Research Center for Complexity Sciences}, \orgname{Hangzhou Normal University}, \orgaddress{\city{Hangzhou}, \postcode{311121}, \state{Zhejiang}, \country{P. R. China}}}
\affil[2]{\orgdiv{College of Media and International Culture}, \orgname{Zhejiang University}, \orgaddress{\city{Hangzhou}, \postcode{310058}, \state{Zhejiang}, \country{P. R. China}}}
\affil[3]{\orgdiv{Gilbert S.\,Omenn Department of Computational Medicine and Bioinformatics}, \orgname{University of Michigan}, \orgaddress{\city{Ann Arbor}, \postcode{48109-2218}, \state{MI}, \country{USA}}}
\affil[4]{\orgdiv{Department of Mathematics}, \orgname{University of Michigan}, \orgaddress{\city{Ann Arbor}, \postcode{48109-1043}, \state{MI}, \country{USA}}}
\affil[5]{\orgdiv{Center for Computational Social Science}, \orgname{Kobe University}, \orgaddress{\city{Kobe}, \postcode{657-8501}, \state{Hyogo}, \country{Japan}}}
\affil[6]{\orgdiv{Complex System Research Center}, \orgname{Shanxi University}, \orgaddress{\city{Taiyuan}, \postcode{030006}, \state{Shanxi}, \country{P. R. China}}}


\abstract{Source detection is crucial for capturing the dynamics of real-world infectious diseases and informing effective containment strategies. Most existing approaches to source detection focus on conventional pairwise networks, whereas recent efforts on both mathematical modeling and analysis of contact data suggest that higher-order (e.g., group) interactions among individuals may both account for a large fraction of infection events and change our understanding of how epidemic spreading proceeds in empirical populations. 
In the present study, we propose a message-passing algorithm, called the HDMPN, for source detection for a stochastic susceptible-infectious dynamics on hypergraphs. By modulating the likelihood maximization method by the fraction of infectious neighbors, HDMPN aims to capture the influence of higher-order structures and do better than the conventional likelihood maximization. We numerically show that, in most cases, HDMPN outperforms benchmarks including the likelihood maximization method without modification.}

\keywords{Hypergraph, Source detection, Epidemic processes, Dynamic message passing}



\maketitle

\section{Introduction}\label{Intro}

Population structure has a large impact on how infectious diseases spread in communities and on the globe \citep{keeling2008modeling,kiss2017mathematics,lu2021data,birello2024estimates}. With the rapid expansion of data on contact networks, identifying the origin of epidemic outbreaks from data has become a pressing research challenge~\citep{wang2020covid,peng2023nlsi}.  The earliest spreaders often constitute a small subset of nodes that strongly drive transmission dynamics~\citep{antulov2015identification,wang2017unification}. Therefore, their timely identification is expected to facilitate targeted interventions such as vaccination, quarantine, or awareness campaigns to be strategically deployed, thereby disrupting transmission chains and reducing the risk of large-scale outbreaks~\citep{pinto2012locating,comin2011identifying, kitsak2010identification}.

Identifying the source node of a spreading process in networks is widely recognized as an NP-hard problem~\citep{brightwell1991counting, valiant1979complexity}. Source detection remains an active and evolving area, with researchers striving to improve and balance its accuracy and computational efficiency. Although source detection has been extensively studied over the past decade, most existing methods are limited to conventional networks that consider only pairwise interactions between nodes. However, real-world populations often exhibit group-based interactions rather than merely dyadic ones, and the former can be captured by higher-order population structures~\citep{battiston2020networks,lambiotte2019networks,bianconi2021higher}. Such higher-order structures can significantly alter diffusion dynamics.  For example, hypergraphs have been reported to accelerate the spreading of infection or reduce the epidemic threshold~\citep{burgio2024triadic,liu2023epidemic}. Hypergraphs can also lead to discontinuous (i.e., first-order) phase transitions marking the onset of epidemic outbreak, in contrast to the continuous transitions (i.e., second-order phase transitions) typically observed in conventional networks~\citep{iacopini2019simplicial,battiston2020networks}. These previous results motivate us to build source detection methods for epidemic processes on hypergraphs (and other higher-order structures such as simplicial complexes). It is not trivial whether direct extensions of existing algorithms originally designed for conventional networks perform similarly well for hypergraphs.

The problem of source detection in hypergraphs remains underexplored. To address this problem, Yu et al.~\citep{yu2024source} proposed a dynamic message passing (DMP) algorithm constructed on a factor graph, where both nodes and hyperedges of the original hypergraph are represented as nodes of a bipartite graph; the hypergraph and its corresponding factor graph are mathematically equivalent. However, the adopted spreading dynamics, modeled via metapaths that traverse from a node to a hyperedge and then to another node, deviate from the actual diffusion mechanisms observed in real-world higher-order systems, where infection occurs directly between nodes. To better solve the source detection problem in real-world systems, we propose an algorithm referred to as the higher-order dynamic message passing modulated with neighbor infection probability (HDMPN), which infers the source node of infection under a hypergraph susceptible-infectious (SI) epidemic process model~\citep{suo2018information}. Central to the HDMPN is the use of the higher-order neighbor infection probability, which is a factor heuristically defined to take advantage of the connectivity and local infection patterns in hypergraphs with the aim of enhancing the accuracy of inference. We demonstrate with numerical experiments on synthetic and empirical hypergraphs that the HDMPN algorithm consistently outperforms baselines. 

The remainder of this paper is organized as follows. Section~\ref{sec: Preliminaries} introduces the main notation and definitions used in the remainder of the paper, along with the SI dynamics model. Section \ref{Algorithms} explains the HDMPN algorithm. Section~\ref{sec:Experiment} presents the baseline methods and provides numerical results. Finally, Section~\ref{conclusion} concludes the paper.

\section{Preliminaries}\label{sec: Preliminaries}

In this section, we introduce the hypergraph and describe the stochastic contagion process used in this study.

\subsection{Hypergraph and notations}

\begin{figure}[htp]
    \centering
    \includegraphics[width=1\linewidth]{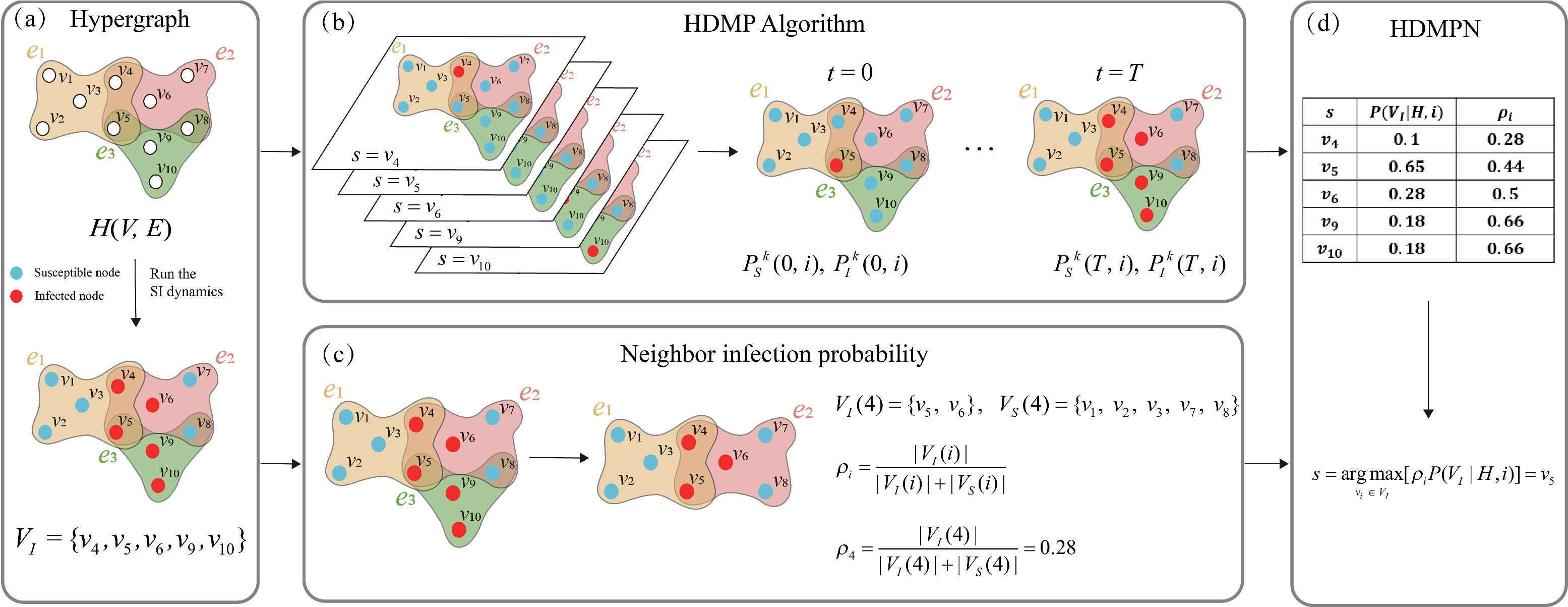}
    \caption{Schematic of the HDMPN algorithm. (a) Final infection snapshot generated by a hypergraph-based susceptible-infected (SI) model at time $T$. (b) Likelihood estimation for each candidate source node using the infection status of all the nodes. The HDMP algorithm selects the candidate node that maximizes the estimated likelihood. (c) Computation of the neighbor infection probability for each candidate node. (d) Integration of the likelihood and neighbor infection probability to infer the most probable source node.}
    \label{fig:schem}
\end{figure}

We denote a hypergraph by $H=(V, E)$, where $V=\{v_1, v_2, \cdots, v_N\}$ is the set of nodes with $N$ being the number of nodes, and $E=\{e_1, e_2, \cdots, e_M\}$ is the set of hyperedges with $M$ being the number of hyperedges. See the upper part of Figure~\ref{fig:schem}(a) for an example. We summarize the primary notations used throughout this paper in Table \ref{tab:notation}. Figure~\ref{fig:schem} as a whole shows the overall procedures of HDMPN.

\begin{table}[ht]
    \centering
    \caption{Main notations used in the present article.}
    \label{tab:notation}
    \renewcommand{\arraystretch}{1.2} 
    \begin{tabular}{p{0.25\linewidth}@{\hspace{10pt}}p{0.7\linewidth}}
        \hline
        Notation & Definition \\ \hline
        $\lambda$ & Probability of contagion  \\
        $H$ & Hypergraph \\
        $T$ & Final time of the SI dynamics\\
        $V_S$ & Set of susceptible nodes at time $T$\\
        $V_I$ & Set of infectious nodes at time $T$\\
        $\left| \Gamma_S(i) \right|$ & Number of susceptible neighbors of $v_i$ at time $T$\\
        $\left| \Gamma_I(i) \right|$ & Number of infectious neighbors of $v_i$ at time $T$\\
        $P_S^k(t, i)$ & Probability that $v_k$ is susceptible at time $t$ when $v_i$ is the source node\\
        $P_I^k(t, i)$ & Probability that $v_k$ is infectious at time $t$ when $v_i$ is the source node\\
        $P^i_S(t)$ & Probability that $v_i$ is susceptible at time $t$  \\
        $P^i_I(t)$ & Probability that $v_i$ is infectious at time $t$  \\
        $D_i$ & Event in which $v_i$ remains susceptible until time $T$\\
        $D_{ij}$ & Event in which adjacent nodes $v_i$ and $v_j$ remain susceptible until time $T$\\
        $d^{k \to i}(t)$ & Infection signal from node $v_k$ to $v_i$ at time $t$  \\
        $\theta^{k\centernot\to i}(t)$ & Probability that $v_k$ has not infected $v_i$ until time $t$ conditioned on event $D_i$  \\
        $\phi^{k\centernot\to i}(t)$ & Probability that $v_k$ is infectious at time $t$ and it has not infected $v_i$ until time $t$ conditioned on event $D_i$\\
        $q_i(t)$ & State (i.e., susceptible or infectious) of $v_i$ at time $t$\\
        $\partial i$ & Set of nodes that share at least one hyperedge with $v_i$\\
        $\partial i \setminus j$ & Set of nodes that share at least one hyperedge with $v_i$, excluding $v_j$\\
        $\eta _{ij}$ & Proportion of the number of hyperedges shared by $v_i$ and $v_j$ to the number of hyperedges to which $v_j$ belongs\\
        \hline
    \end{tabular}
\end{table}

\subsection{Contagion dynamics model}

We use a discrete-time stochastic susceptible-infectious (SI) model that incorporates higher-order interactions. At any time, the nodes are in one of the two states, i.e., susceptible (S) or infectious (I)~\citep{xie2023efficient, suo2018information}; see the lower panel of Figure~\ref{fig:schem}(a).
The model runs as follows.
\begin{itemize}
\item At $t=0$, the process is initialized with one infectious seed node, while all other $ N-1$ nodes are susceptible.
\item At each discrete time $t$, every node $v$ infectious at time $t-1$ selects one of its associated hyperedges $e$ uniformly at random. Each susceptible node within $e$ then independently becomes infected by $v$ with probability $\lambda$. The eventually obtained state of the nodes is the state at time $t$.
\item We increment $t$ by one and repeats the infection dynamics. The process stops at a final time $T$.
\end{itemize}

\section{Higher-order dynamic message-passing modulated with neighbor infection probability (HDMPN)}\label{Algorithms}
We focus on the problem of detecting a single source node in a hypergraph, given the state (i.e., susceptible or infectious) of all nodes at time $T$. To this end, we propose the HDMPN algorithm to estimate the likelihood of each node being the source of the SI process.
Let $V_I$ denote the set of infectious nodes at time $T$. By leveraging Bayesian inference, we formulate the source detection task as a maximum likelihood estimation problem as follows:
\begin{equation}
    v_{\hat{i}}= \underset{v_i\in V_I}{\text{argmax}}\quad \text{Prob}(i|H, V_I) = \underset{v_i\in V_I}{\text{argmax}}\quad \text{Prob}(V_I|H, i),\label{eq1}
\end{equation}
where $\text{Prob}(\cdot | \cdot)$ represents the conditional probability. For example, $\text{Prob}(V_I|H,i)$ is the probability that, given $v_i$ as the source of transmission, the disease propagates in the hypergraph to produce the set of infectious nodes, $V_I$, at time $T$.
We refer to the maximization given by Eq.~\eqref{eq1} as the higher-order dynamic message-passing algorithm (HDMP) and use it as a baseline algorithm in our numerical simulations. See Figure~\ref{fig:schem}(b) for a schematic.

Conditioned that $v_i$ is the source node, we denote the probability that node $v_j$ is susceptible or infectious at time $t$
by $P^j_S(t, i)$ and $P^j_I(t, i)$, respectively. Under the approximation that the states of the nodes except $v_i$ are independent of each other, we can compute $\text{Prob}(V_I |H, i)$ as the product of the probabilities that each node is either susceptible or infectious. Accordingly, we write
\begin{equation}
    \text{Prob}\left( V_I | H, i  \right) =\prod_{v_k\in V_I }^{}P^{k}_{I}\left( T, i \right)   \prod_{v_l\in V_S }^{}P^{l}_{S}\left( T, i \right) ,    \label{eq2n}
\end{equation}
where $V_S$ represents the set of susceptible nodes at time $T$. 

Now, to explain the HDMPN algorithm, we note that, for an infectious node, a higher infection probability among its neighboring nodes may increase the likelihood of it being the initial source node. To capture this intuition, we introduce the concept of neighbor infection probability, which quantifies the propensity that the neighbors of a given node are also infected at the final time $T$. We define the neighbor infection probability of node $v_i$ as
\begin{equation}
    \rho _i = \frac{\left | \Gamma_S(i) \right | }{\left | \Gamma_I(i) \right |  + \left |  \Gamma_S(i)\right | }  ,
    \label{eq3}
\end{equation}
where $\left | \Gamma_S(i) \right |$ and $\left |  \Gamma_I(i)\right |$ represent the number of susceptible and infectious neighbors of $v_i$ at time $T$, respectively. See Figure~\ref{fig:schem}(c) for an example.
We propose a source detection algorithm for the hypergraph SI model, i.e., HDMPN algorithm, that integrates the neighbor infection probability with the likelihood of node $v_i$ being the source. The HDMPN method is defined by
\begin{equation}
    \underset{v_i\in V_I}{\text{argmax}}\quad \left[\rho _i \text{Prob}(V_I|H, i)\right].
    \label{eq2}
\end{equation}
See Figure~\ref{fig:schem}(d) for an example.
In the following text, we provide a detailed derivation of $\text{Prob}(V_I \mid H, i)$.

We introduce two auxiliary events. For any node $v_i$ that is not initially infectious, $D_i$ denotes the event in which $v_i$ remains susceptible until time $T$. Similarly, $D_{ij}$ represents the event in which a pair of adjacent nodes, i.e., $v_i$ and $v_j$, remain susceptible until time $T$. Under event $D_i$, no infection originates from $v_i$, allowing us to assume that the neighboring branches of $v_i$ are independent. This assumption implicitly requires the hypergraph to be locally tree-like, which ensures that short loops are rare.

To derive the message-passing equations, we further introduce three auxiliary variables. First, let $\theta^{k\centernot\to i} (t)$ represent the probability that infection has not transmitted from node $v_k$ to node $v_i$ by time $t$ under event $D_i$. In other words, we write
\begin{equation}
    \theta ^{k\centernot\to i} (t)=\text{Prob}\left(\sum_{t'=1}^{t}d^{k\to i}(t')=0\biggm| D_i\right),
    \label{theta}
\end{equation}
where $d^{k\to i}(t)$ denotes the infection signal, which is equal to $1$ if $v_k$ infects $v_i$ at time $t$, and 0 otherwise. Next, we define $\phi ^{k\centernot\to i}(t)$ as the probability that $v_k$ is infectious at time $t$ and $v_k$ has not infected $v_i$ by time $t$ under event $D_i$. In other words, we write
\begin{equation}
    \phi ^{k\centernot\to i}(t)=\text{Prob}\left ( \sum_{t'=1}^{t}d^{k\to i}(t')=0, q_k(t)=I \biggm| D_i \right ),
    \label{phi}
\end{equation}
where $q_k(t)=I$ indicates the event that node $v_k$ is infectious at time $t$. We also write $q_k(t)=S$ when $v_k$ is susceptible at time $t$.
\begin{lemma}
Under the assumption that the neighboring nodes of $v_i$ act independently, it holds true that
\begin{equation}
\text{Prob}\left(q_i(t) = S \mid D_{j}\right) = P_S^i(0) \prod_{v_k \in \partial i \setminus j} \theta^{k \centernot\to i}(t),
\label{eq:lemma1-statement}
\end{equation}
where $P_{S}^{i} (t)$ is the probability that node $v_i$ is susceptible at time $t$, and $\partial i \setminus j$ represents the set of nodes that share at least one hyperedge with node $v_i$, excluding $v_j$.
\end{lemma}
\begin{proof}
The probability $\text{Prob}\left(q_i(t) = S \mid D_{j}\right)$ depends on the state of node $v_i$ at time $t=0$, denoted as $P_S^i(0)$, as well as the cumulative influence exerted by its neighbors over time. To ensure that $q_i(t) = S$, it is sufficient that all neighbors of $v_i$, except for $v_j$, fail to infect $v_i$ prior to time $t$. Therefore, we obtain
\begin{equation}
    \text{Prob}\left(q_i(t) = S \mid D_{j}\right) = P^i_S(0)\text{Prob}\left(\sum_{k\in\partial i\setminus j}\sum_{t^{\prime}=1}^{t}d^{k\to i}(t^{\prime})\biggm|D_j\right),
    \label{proof1}
\end{equation}
where we recall that $d^{k \to i}(t')$ denotes the infection signal transmitted from neighbor $v_k$ to node $v_i$ at time $t'$. Moreover, under the condition that $q_i(t) = S$, the event $D_j$ is equivalent to $D_{ij}$. Therefore, we can rewrite Eq.~(\ref{proof1}) as
\begin{equation}
    \text{Prob}\left(q_i(t) = S \mid D_{j}\right) = P^i_S(0)\text{Prob}\left(\sum_{k\in\partial i\setminus j}\sum_{t^{\prime}=1}^{t}d^{k\to i}(t^{\prime})\biggm|D_{ij}\right).
    \label{proof2}
\end{equation}
The double summation in Eq.~(\ref{proof2}) only involves infection signals sent to $v_i$ from its neighbors excluding $v_j$. Under the locally tree-like assumption, the state of $v_j$ at any time does not influence this transmission event. Therefore, the condition $D_{ij}$ can be simplified to $D_i$ in Eq.~\eqref{proof2}, allowing us to rewrite Eq.~(\ref{proof2}) as
\begin{equation}
    \text{Prob}\left(q_i(t) = S \mid D_{j}\right) =
    P_S^i(0)\prod_{k\in\partial i\setminus j}\mathrm{Prob}\left(\sum_{t^{\prime}=1}^{t} d^{k\to i}(t^{\prime})\biggm| D_i  \right).
    \label{proof3}
\end{equation}
By substituting Eq.~(\ref{theta}) in Eq.~(\ref{proof3}), we obtain Eq.~(\ref{eq:lemma1-statement}).
\end{proof}
In terms of $\theta^{k\centernot\to i}(t)$, $\phi^{k \centernot\to i}(t)$ and $\text{Prob}\left(q_i(t) = S \mid D_{j}\right)$, we now explain the recursive rules for computing the probability $P^{i}_S(t+1, \hat{s})$, where we remind that $\hat{s}$ is the assumed source node. For simplicity, we abbreviate $P^{i}_S(t+1, \hat{s})$ to $P^{i}_S(t+1)$ in the following derivations. All other variables and parameters are likewise defined under the assumption that $\hat{s}$ is the source node being evaluated.

According to the recursive relationship, node $v_i$ remains susceptible at time $t+1$ only if all of its potentially infectious neighbors fail to infect $v_i$ by time $t+1$. Therefore, we obtain
\begin{equation}
    P^{i}_S(t+1)=P^i_S(0)\prod _{v_k\in \partial i}\theta ^{k\centernot\to i}(t+1),
    \label{eq13}
\end{equation}
where $\partial i$ represents the set of nodes that share at least one hyperedge with $v_i$.

Next, to compute $\theta^{k \centernot\to i}(t + 1)$, we note that the difference between $\theta^{k \centernot\to i}(t + 1)$ and $\theta^{k \centernot\to i}(t)$ is the probability that $v_k$ infects $v_i$ at time $t$. This probability is given by
\begin{equation}
    \theta ^{k\centernot\to i}(t+1)-\theta ^{k\centernot\to i}(t)=-\phi^{k\centernot\to i}(t)\lambda \eta _{ki},
    \label{eq15}
\end{equation}
where $\eta_{ki}$ is the fraction of hyperedges shared by $v_k$ and $v_i$ among the hyperedges to which $v_k$ belongs. Infectious node $v_k$ selects one of the hyperedges containing $v_i$ with probability $\eta_{ki}$ for potentially infecting $v_i$. We remind that $\lambda$ is the probability that $v_k$ infects $v_i$ once such a shared hyperedge is selected and that $\phi^{k \centernot\to i}(t)$ is the probability that $v_k$ is infectious and $v_i$ remains susceptible at time $t$. We compute $\phi^{k \centernot\to i}(t)$ as follows:
\begin{equation}
\begin{split}
     \phi^{k\centernot\to i}(t) & =\phi^{k\centernot\to i}(t-1)-\phi^{k\centernot\to i}(t-1)\lambda \eta _{ki}\\
     &+ \text{Prob}\left(q_k(t - 1)= S \mid D_{i}\right)-\text{Prob}\left(q_k(t) = S \mid D_{i}\right).\label{eq16}
\end{split}
\end{equation}
The left-hand side of Eq.~\eqref{eq16} captures the probability that $v_k$ has the potential to infect $v_i$ but the transmission does not occur at time $t$. The difference between $\phi^{k \centernot\to i}(t)$ and $\phi^{k \centernot\to i}(t - 1)$ originates from two scenarios: (\romannumeral 1) $v_k$ infects $v_i$ at time $t-1$, which the second term on the right-hand side of Eq.~\eqref{eq16} represents; and (\romannumeral 2) $v_k$ contracts infection at time $t-1$, which $\text{Prob}\left(q_k(t - 1) = S \mid D_{i}\right)-\text{Prob}\left(q_k(t) = S \mid D_{i}\right)$ represents. Consequently, the problem reduces to computing $\text{Prob}\left(q_k(t) = S \mid D_{i}\right)$. Lemma 1 implies that
\begin{equation}
    \text{Prob}\left(q_k(t) = S \mid D_{i}\right)=P^i_S(0)\prod _{v_j\in \partial i\setminus k}\theta ^{j\centernot\to i}(t).\label{eq17}
\end{equation}
Therefore, variables $\theta^{i \centernot\to j}(t+1)$, $\phi^{i \centernot\to j}(t+1)$ and $\text{Prob}\left(q_i(t+1) = S \mid D_{j}\right)$ can be recursively expressed in terms of these variables at time
$t$.

The initial condition of this recursive computation process is given by
\begin{equation}
    \text{Prob}\left(q_i(0) = S \mid D_{j}\right) = 
    \begin{cases} 
        0 & \text{if } q_i(0) = 1 \text{ or } q_j(0) = 1, \\
        1 & \text{otherwise},
        \label{eq20}
    \end{cases}
\end{equation} 
\begin{equation}
    \theta^{i \centernot\to j}(0) = 1,\label{eq18}
\end{equation}
and
\begin{equation}
    \phi^{i \centernot\to j}(0) = 
    \begin{cases} 
        1 & \text{if } q_i(0) = 1, \\
        0 & \text{otherwise}.
    \end{cases}\label{eq19}
\end{equation}
Using the recursive rules given by Eqs.~\eqref{eq13}--\eqref{eq19}, we first compute all relevant probability variables for each node iteratively from $t=1$ to $t=T$. Then, using Eq.~\eqref{eq2n} with $P_I^i(T) = 1 - P_S^i(T)$, we calculate the likelihood $P(V_I|H, i)$ of observing the finally infected nodes, $V_I$, given hypergraph $H$ and the source node $v_i$ (see Figure~\ref{fig:schem}(b)). Then, for HDMP and HDMPN, we exhaustively look for the node realizing the maximization given by Eq.~\eqref{eq1} and Eq.~\eqref{eq2}, respectively, and use the maximizer as the estimated source node. See Figure~\ref{fig:schem}(d) for visual explanation of this procedure for HDMPN.

The time complexity of the HDMP and HDMPN algorithms is as follows. The HDMP algorithm runs in $O(N^2 T)$ time. This is because the message-passing variables, such as $\theta^{k \centernot\to i}(t)$ and $\phi^{k \centernot\to i}(t)$ (see Eq.~\eqref{theta} and Eq.~\eqref{phi}), need to be updated for all $O(N^2)$ node pairs at each of the $T$ time steps. The HDMPN algorithm additionally requires a one-time calculation of the neighbor infection probability $\rho_i$ for each node, which introduces an additional $O(N)$ time. Therefore, the total complexity of HDMPN is $O(N^2T + N)$, which reduces to $O(N^2 T)$.

\section{Experimental results}\label{sec:Experiment}

In this section, we introduce the baseline algorithms, explain the evaluation metrics, and report the numerical results obtained on synthetic and empirical hypergraphs.

\subsection{Baseline algorithms}

To validate the effectiveness of HDMPN, we introduce baseline methods for identifying the spreading source in the SI model. We use HDMP as a baseline algorithm. To the best of our knowledge, no dedicated methods have been developed specifically for source localization in hypergraphs. Therefore, we adapt three existing baseline methods for the SI dynamics on conventional networks to the case of hypergraphs. For each baseline method, ties may occur when multiple nodes have exactly the same highest score. In such cases, the predicted source node is one of the tying nodes selected uniformly at random.

\textbf{Higher-order closeness centrality (HCC)}. {HCC generalizes the concept of closeness centrality from conventional networks to hypergraphs. The underlying intuition is that nodes with shorter average distances to all other nodes in the hypergraph composed only of the finally infectious nodes are more likely to be the infection source. Formally, the HCC of an infectious node is defined as the reciprocal of the sum of its shortest-path distances to all other infectious nodes in the so-called infected hypergraph, $\overline{H}^I$. By definition, the node set of $\overline{H}^I$ is $V_I$, i.e., the infectious nodes at time $T$. For each hyperedge of $H$, denoted by $e$, the subset of $e$ that is composed of the nodes in $V_I$ is a hyperedge of  $\overline{H}^I$. For example, suppose that $e = \{v_1, v_2, v_3\}$ is a hyperedge in the original hypergraph. Then, if $v_1$ and $v_2$ are infectious and node $v_3$ is still susceptible at time $T$, $e' = \{v_1, v_2\}$ is a hyperedge of $\overline{H}^I$. Isolated nodes are excluded from $\overline{H}^I$. If a hyperedge $e'$ appears multiple times, we count its occurrences with multiplicity.

Let $n$ denote the number of nodes in the infected hypergraph $\overline{H}^I$.
To define HCC, we first introduce the concept of $s$-distance~\citep{aksoy2020hypernetwork}. Two hyperedges are defined as $s$-adjacent if they share at least $s$ nodes. We define an $s$-walk of length 
$l$ connecting hyperedges $e_{i_0}$ and $e_{i_l}$ as a sequence of hyperedges:
\begin{equation}
    w^l_i = \{e_{i_0},e_{i_1}, \ldots, e_{i_{l - 1}},e_{i_l}\},
\end{equation}
where each consecutive pair of hyperedges is $s$-adjacent. An $s$-path between two hyperedges $e_{i_0}$ and $e_{i_l}$ is defined as a sequence of distinct hyperedges, ensuring that no hyperedge appears more than once in the $s$-walk. The shortest $s$-distance $d^e_s(p, q)$ between two hyperedges $e_p$ and $e_q$ is defined as the length of the shortest $s$-path connecting them. If no such $s$-path exists, their $s$-distance is defined to be $\infty$. Building upon this definition of hyperedge distance, we define the $s$-distance between nodes. Let $E_i$ denote the set of hyperedges that include node $v_i$. We define the distance between nodes $v_i$ and $v_j$, denoted by $d^v_s(i, j)$, by
\begin{equation}\label{NodeDistance}
    d_{s}^{v}(i, j) = 
    \begin{cases}
        0 & \text{if } i = j,\\
        1 & \text{if } i \neq j \text{ and } E_i \cap E_j \neq \emptyset, \\
        \min\limits_{e_p \in E_i, e_q \in E_j}d_{s}^{e}(p, q)\ + 1 & \text{otherwise.}
    \end{cases}
\end{equation}

Finally, the HCC of node $v_i$, denoted by $C_C^H(i)$, is given by
\begin{equation}\label{HDC}
    C_C^H(i)=\sum_{s = 1}^{s_{\max}}\alpha ^s\sum_{j\ne i}^{n}\frac{1}{d^v_s(i, j)},
\end{equation}
where $s_{\max}$ is a parameter indicating the maximum value of the $s$-distance we consider, and $\alpha$ is a parameter controlling the contribution of different $s$-distances to $C_C^H(i)$. 

\textbf{Higher-order betweenness centrality (HBC)}.
The HBC extends the betweenness centrality for conventional networks to the case of hypergraphs. Specifically, we identify all $s$-paths in the infected hypergraph $\overline{H}^I$ at time $T$. The HBC of node $v_i$, denoted by $C_B^H(i)$, is defined by
\begin{equation}
    C_B^H(i) = \sum_{s=1}^{s_{\max}} \alpha^s \sum_{q \neq v_i \neq u} \frac{\sigma_{qu}^s(v_i)}{\sigma_{qu}^s},
\end{equation}
where $\sigma_{qu}^s$ denotes the number of shortest $s$-paths between nodes $v_q$ and $v_u$, and $\sigma_{qu}^s(v_i)$ represents the number of such shortest paths that pass through $v_i$. 

\textbf{Higher-order Monte Carlo-based soft boundary estimation method (HMCSM)}.
In addition to the HCC and HBC, which are centrality-based methods, the infection source can also be inferred using simulation-based approaches, such as the Monte Carlo-based soft boundary estimation method (MCSM)~\citep{antulov2015identification}. Here, we extend MCSM to the case of hypergraphs, proposing the higher-order MCSM (HMCSM). For each node $v_i$ in $V_I$, where we remind that $V_I$ is the set of finally infectious nodes, we simulate the SI model on the hypergraph $H$ for $T$ time steps.  This process generates a set of finally infectious nodes, denoted by ${V}_I'$, for each candidate source node $v_i \in V_I$. To evaluate the likelihood of each node being the actual source, we compute the Jaccard similarity between $V_I$ and ${V}_I'$, i.e.,
$J(V_I, {V}_I') \equiv \frac{ |V_I \cap {V}_I'|} {|V_I \cup {V}_I'|}$.
We then select the node with the highest Jaccard similarity as the most probable infection source. 
To mitigate the impact of stochasticity of the spreading dynamics, we run the SI model $R=100$ times starting from each $v_i \in V_I$. We define a score for each node $v_i$ by
\begin{equation}
    P_i = \frac{1}{R}\sum_{r=1}^{R} \exp \left[  -(1-J \left (V_I, {V}_I'(i,r)  \right ))^2 \right],
    \label{HMSCM}
\end{equation}
where ${V}_I'(i, r)$ is the set of the eventually infected nodes in the $r$th run when the infection source is $v_i$. 
The HMCSM uses the node with the largest $P_i$ as the predicted infection source.

\subsection{Hypergraphs used in the numerical simulations}

We use the following synthetic and empirical hypergraphs; see Table \ref{infor} for their properties, including the dependence on the model parameters.

We use the following three hypergraph generation models. 
\begin{itemize}
    \item \textbf{Erd\H{o}s-R\'{e}nyi Hypergraph model (ERH)} \citep{surana2022hypergraph}. The ERH constructs a $k^E$-uniform hypergraph (i.e., each hyperedge is composed of $k^E$ nodes) $H=(V, E)$ by randomly selecting nodes to form hyperedges, in which each hyperedge contains exactly $k^E$ nodes. We set $k^E = 5$. Initially, the hyperedge set $E$ is empty. We add each hyperedge $e$ one by one, where $e$ is the set of $k^E$ distinct nodes that are selected from the $N$ nodes uniformly at random. 
    If the selected combination of nodes does not exist in the current set of hyperedges, $E$, we add it to $E$; otherwise, we discard the combination and retry. We repeat this process until we obtain $M$ unique hyperedges.

    \item \textbf{Scale-free hypergraph model (SFH)} \citep{feng2024hyper}. The SFH model constructs a $k^E$-uniform hypergraph in which the node hyperdegree follows a power-law distribution 
    $p(k^H)\propto (k^H)^{-\gamma}$, where $\gamma$ is a parameter. We set $k^E=5$ and $\gamma=3$. The model first generates the expected hyperdegree for each of the $N$ nodes, $v_i$, by drawing a random number $k_i^E$ independently obeying the distribution $p(k^H)$ and assigns $v_i$ a probability $p_i = \frac{k^H_i}{ {\textstyle \sum_{j = 1}^{N}}k^H_j }$ of being included in a hyperedge. To generate each hyperedge $e$, we iteratively add non-duplicate nodes based on their sampling probabilities $p_i$ until $k^E$ nodes have been added. If the resulting hyperedge $e$ does not exist in the hypergraph, we add it to $E$. Otherwise, we discard the generated $e$. We repeat this process until we obtain $M$ unique hyperedges.
    
    \item  \textbf{HyperCL model (HCL)} \citep{xie2023efficient,nakajima2021randomizing}. In contrast to the ERH and SRH models, the HCL model generates hypergraphs with non-uniform hyperedge sizes. Given a total of $N$ nodes and $M$ hyperedges, we first generate the expected hyperdegree of each node, $\left\{k_{1}^{H}, k_{2}^{H}, \cdots, k_{N}^{H}\right\}$, according to the hyperdegree distribution $p(k^H) \propto (k^H)^{-\gamma}$, where $\gamma$ is a parameter. We also draw the hyperedge size of each hyperedge, 
$\left\{k_{1}^{E}, k_{2}^{E}, \ldots, k_{M}^{E}\right\}$, 
from a uniform distribution on $\{2, 3, 4, 5 \}$. To generate each hyperedge $e$, we iteratively add nodes to $e$ based on their hyperdegree, with the probability of selecting node $v_i$ given by 
$\frac{k^H_i}{ {\textstyle \sum_{\ell=1}^{N}k^H_{\ell}} }$, until 
the hyperedge reaches its designated size. If the generated hyperedge $e$ does not exist in the hypergraph, we add $e$ to $E$; we discard the generated $e$ otherwise. We repeat this process until we obtain $M$ unique hyperedges.
\end{itemize}

We also use the following three empirical hypergraphs from different domains.
\begin{itemize}
    \item  \textbf{Cat-edge-algebra-questions (Algebra)}~\citep{amburg2020fair} hypergraph is composed of user interactions on the MathOverflow platform, specifically focusing on discussions related to algebra. It comprises user-generated contents in the form of comments, questions, and answers. Each node represents an individual user. Each hyperedge contains the users who engage with the same algebra-related question.
    \item  \textbf{Restaurant-Rev}~\citep{amburg2020fair} hypergraph represents user review activity on Yelp, focusing on restaurant reviews over a one-month period. Each node represents a Yelp user. Each hyperedge contains the users who have reviewed the same restaurant. 
    \item  \textbf{NDC-classes}~\citep{Benson-2018-simplicial} hypergraph represents the components of drugs, with individual substances represented as nodes. A hyperedge represents the set of substances composing a drug.
\end{itemize}

\begin{table}[htp]
    \centering
    \caption{Structural properties of the synthetic and empirical hypergraphs. Here, $N$ and $M$ are the number of nodes and the number of hyperedges, respectively; $\left \langle k \right \rangle $ is the average of the number of a node's neighbors; $\left \langle k^H \right \rangle $ is the average hyperdegree of the node; $\left \langle k^E \right \rangle $ is the average size of the hyperedge. Synthetic hypergraphs are denoted as ERH-$N$, SFH-$N$, and HCL-$\gamma$-$N$, where $N$ indicates the number of nodes and $\gamma$ (for HCL only) specifies the power-law exponent of the hyperdegree distribution.}
    \label{infor}
    \begin{tabular}{c@{\hspace{30pt}}c@{\hspace{30pt}}c@{\hspace{30pt}}c@{\hspace{30pt}}c@{\hspace{30pt}}c@{\hspace{30pt}}c}
        \hline
        Hypergraph & $N$ & $M$ & $\left \langle k \right \rangle $ & $\left \langle k^H \right \rangle $ & $\left \langle k^E \right \rangle $  \\ 
        \hline
        ERH-100 & 100 & 100 & 18.02 & 5.00 & 5.00   \\
        ERH-200 & 200 & 200 & 19.07 & 5.03 & 5.00   \\
        ERH-300 & 300 & 300 & 19.60 & 5.05 & 5.00   \\
        ERH-400 & 400 & 400 & 19.68 & 5.03 & 5.00   \\
        SFH-100 & 100 & 100 & 15.04 & 5.21 & 5.00   \\
        SFH-200 & 200 & 200 & 18.42 & 5.10 & 5.00   \\
        SFH-300 & 300 & 300 & 18.91 & 5.12 & 5.00   \\
        SFH-400 & 400 & 400 & 17.49 & 5.19 & 5.00  \\
        HCL-2.0-100 & 100 & 100 & 8.85 & 4.29 & 3.22   \\
        HCL-2.0-200 & 200 & 200 & 9.15 & 3.85 & 3.04   \\
        HCL-2.0-300 & 300 & 300 & 8.69 & 4.02 & 2.99  \\
        HCL-2.0-400 & 400 & 400 & 9.09 & 4.06 & 2.94   \\
        HCL-2.3-100 & 100 & 100 & 7.47 & 3.27 & 2.81   \\
        HCL-2.3-200 & 200 & 200 & 8.36 & 3.51 & 2.97   \\
        HCL-2.3-300 & 300 & 300 & 8.95 & 3.70 & 3.09   \\
        HCL-2.3-400 & 400 & 400 & 9.16 & 3.66 & 3.03   \\
        HCL-2.5-100 & 100 & 100 & 7.87 & 3.24 & 3.01   \\
        HCL-2.5-200 & 200 & 200 & 8.37 & 3.32 & 2.98  \\
        HCL-2.5-300 & 300 & 300 & 9.03 & 3.49 & 3.03   \\
        HCL-2.5-400 & 400 & 400 & 8.46 & 3.49 & 3.02   \\
        Algebra & 423 & 1268 & 78.90 & 19.53 & 6.52  \\
        Restaurant-Rev & 565 & 601 & 79.75 & 8.14 & 7.66   \\
        NDC-classes & 1161 & 1088 & 10.72 & 5.55 & 5.92   \\
        \hline
    \end{tabular}
\end{table}

\subsection{Numerical simulation setup}

In this section, we explain the parameter settings and performance measures for our numerical experiments. We uniformly randomly select a non-isolated node as the initially infectious node. We run the SI process until it reaches a predefined threshold fraction of infectious nodes. Specifically, for hypergraphs with less than or equal to 1,000 nodes, the SI dynamics stops when more than 10\% of nodes have been infected for the first time. We set the threshold to 5\% for the NDC-classes hypergraph, which has more than 1,000 nodes. Once we have stopped the simulation, we record the time as the final time $T$ and the set of infectious nodes as $V_I$. We calculate the probability that each node is the source using Eqs.~\eqref{eq2n} and \eqref{eq3}. We use these probabilities for each run to compute the performance measures explained below. We repeat each experiment $R=10^3$ times with an independently drawn initially infectious node. Unless we state otherwise, we set the infection probability to $\lambda=0.5$. 

We quantify the performance of the different source detection methods in terms of two quantities, which we refer to as the accuracy and ranking.

Let $s$ be the true source and $\hat{s}$ be the predicted source. We define the accuracy as 
\begin{equation}
     \text{Accuracy} = \frac{R_{\text{true}}}{R}, 
\end{equation}
where $R_{\text{true}}$ denotes the number of runs in which $\hat{s}=s$.

Given a source node $s$ and its corresponding set of finally infectious nodes $V_I$, we denote by $|V_I|$ the number of finally infectious nodes. In each run, for a given source detection method, the nodes are ranked in descending order based on their computed score (i.e., the centrality value for HCC and HBC, the value computed by Eq.~\eqref{HMSCM} for HMCSM, $\text{Prob}(V_I \mid H, i)$ for HDMP, and $\rho_i\text{Prob}(V_I \mid H, i)$ for HDMPN). The position of node $s$ in this ranking is denoted as $\text{pos}(s)$. We define the ranking metric for each run as
\begin{equation}
    \mu(s) = \frac{\text{pos}(s) - 1}{|V_I|}.
\end{equation}
A smaller $\mu(s)$ value indicates better performance, with $\mu(s) = 0$ representing a perfect identification of the source node. We use distributions of $\mu(s)$ calculated on the basis of $R$ independent runs to compare the performance of the different source detection methods.

\subsection{Results}

We compare the performance of the HDMPN algorithm with the baseline algorithms on hypergraphs generated using the ERH model in Figure~\ref{figuretime}. To assess whether HDMP or HDMPN performs better than HCC with different $\alpha$ values, we compute the accuracy for the HCC using the $\alpha$ value that maximizes the accuracy among $\alpha \in [0.1, 3.0]$ with an increment of $0.001$. We do the same for the accuracy of the HBC and the ranking-based performance comparison for the HCC and HBC. Note that the $\alpha$ value that optimizes the accuracy of the HCC and the value that optimizes the ranking metric of the HCC may be different from each other; the same applies for the HBC.
\begin{figure}[hbp]
    \centering
    \includegraphics[width=1\textwidth]{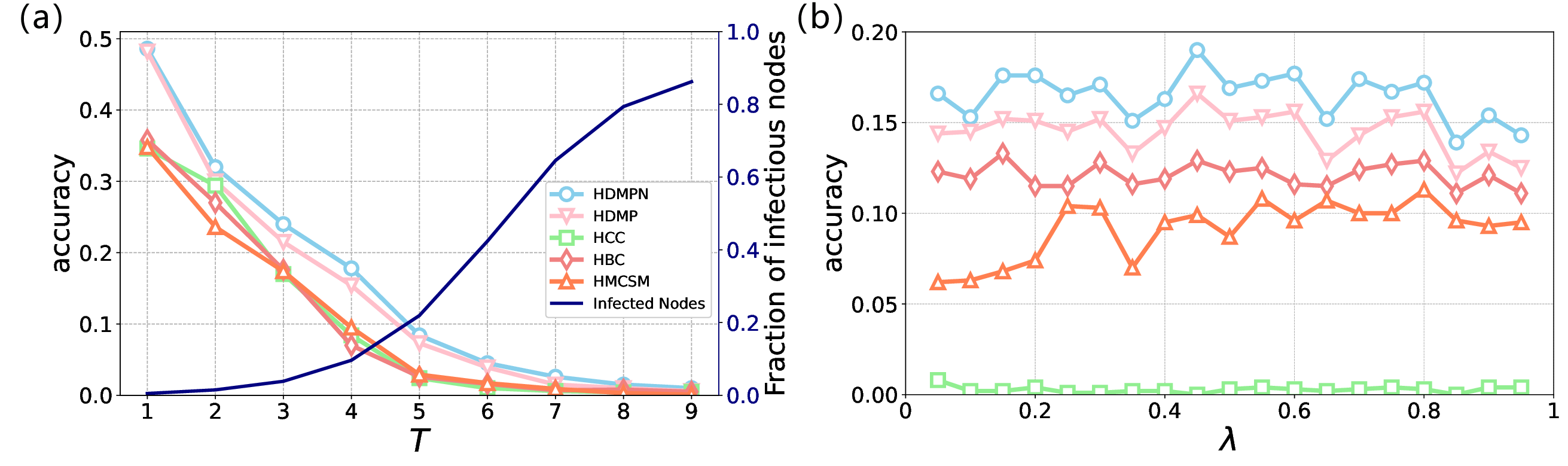}
    \caption{Accuracy of the five source detection algorithms on a hypergraph generated by the ERH with $N = 200$, $M = 200$, and $\langle k^H \rangle = 5$. (a) Accuracy over different simulation times $T$. We set the infection probability to $\lambda = 0.5$.
(b) Accuracy as a function of the infection probability, $\lambda$. In (b), we ran each simulation until at least 10\% of nodes contract infection.}
    \label{figuretime}
\end{figure}
We set $N = 200$, $M = 200$, and an average hyperdegree $\langle k^H \rangle = 5$. We show in Figure~\ref{figuretime}(a) the fraction of infectious nodes and the accuracy of the five source detection methods over a range of $T$. In this figure, $\lambda=0.5$ is held constant.
We find that the accuracy declines as $T$ increases, i.e., as the number of infectious nodes increases, regardless of the source detection method. The accuracy is higher at smaller $T$ intuitively because there are fewer candidates of source nodes (i.e., infectious nodes) at small times. We also find that HDMPN and HDMP consistently outperform the other three baseline methods across the $T$ values and that it is especially the case before approximately 75\% of nodes have been infected. The figure shows that HDMPN slightly outperforms HDMP. Figure~\ref{figuretime}(b) shows the dependency of the accuracy on the infection probability, $\lambda$, when the SI dynamics is stopped once approximately 10\% of the nodes have contracted infection. While the performance of each method fluctuates as $\lambda$ varies, HDMPN is consistently better than all the baseline algorithms, including HDMP.

\begin{figure}[h]
    \centering
    \includegraphics[width=1\textwidth]{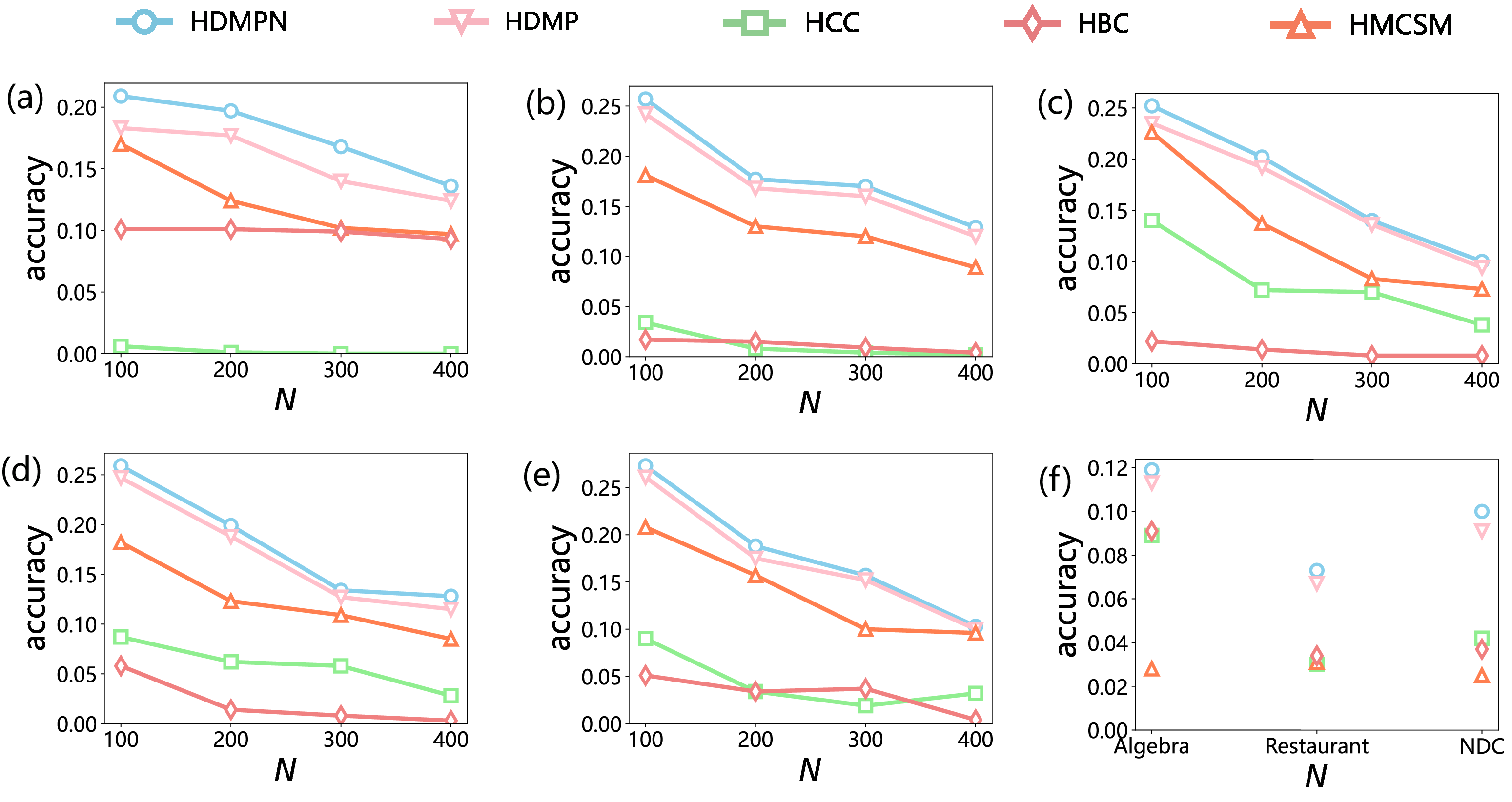}
    \caption{Accuracy of source detection for the five source detection methods and hypergraphs. 
(a) ERH model;
(b) SFH model;
(c-e) HCL model with (c) $\Theta = 2.0$, (d) $\Theta = 2.3$, and (e) $\Theta = 2.5$;
(f) empirical hypergraphs. We set $\lambda = 0.5$ and evaluated the accuracy when at least 10\% of nodes have been infected for all but the NDC-classes hypergraph; we use the threshold of 5\% for the NDC-classes hypergraph because it has substantially more nodes than the other hypergraphs.}
    \label{accuracy}
\end{figure}

We further assess the accuracy of the different source detection methods on synthetic hypergraphs of various sizes.
Specifically, we generate hypergraphs using the ERH, SFH, and HCL models, with the number of nodes $N \in \{100, 200, 300, 400 \}$ and a fixed maximum hyperedge size of 5.
We show in Figure~\ref{accuracy}(a), (b), (c)--(e) the accuracy on the ERH, SFH, and HCL models, respectively.
We find that the accuracy is lower for larger hypergraphs. 
We also find that HDMPN consistently outperforms all baseline methods on all hypergraphs and for all $N$ values, in particular on smaller hypergraphs. For instance, when the hypergraph size is relatively small (e.g., $N = 100$), 
\clearpage
\newpage
\begin{figure}[h]
    \centering
    \includegraphics[width=1\linewidth]{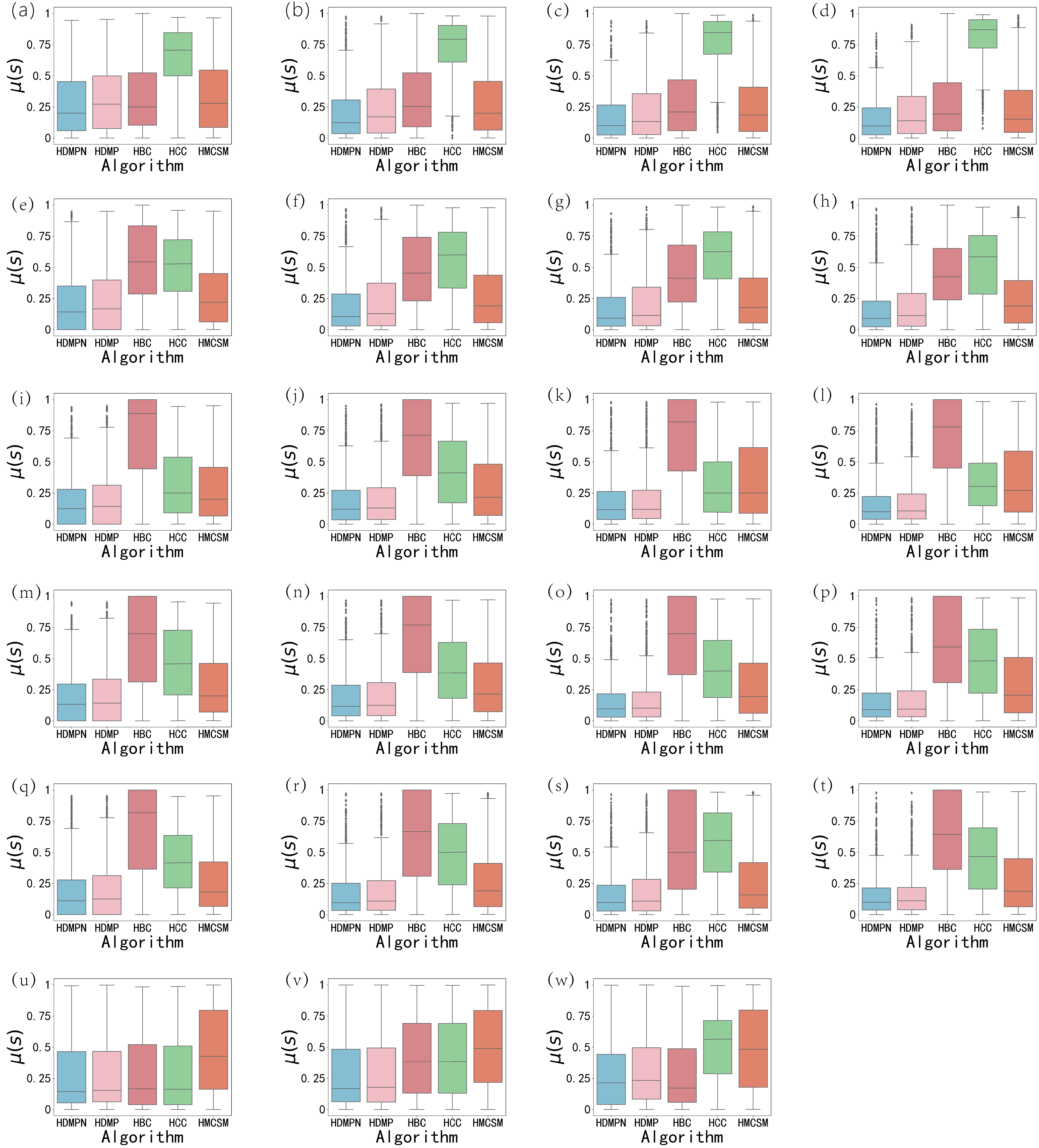}
    \caption{Ranking-based performance of the five source detection methods on synthetic and empirical hypergraphs.
(a) ERH, $N=100$. (b) ERH, $N=200$. (c) ERH, $N=300$. (d) ERH, $N=400$.
(e) SFH, $N=100$. (f) SFH, $N=200$. (g) SFH, $N=300$. (h) SFH, $N=400$.
(i) HCL, $(N, \Theta) = (100, 2.0)$. (j) HCL, $(N, \Theta) = (200, 2.0)$.
(k) HCL, $(N, \Theta) = (300, 2.0)$. (l) HCL, $(N, \Theta) = (40, 2.0)$.
(m) HCL, $(N, \Theta) = (100, 2.3)$. (n) HCL, $(N, \Theta) = (200, 2.3)$.
(o) HCL, $(N, \Theta) = (300, 2.3)$. (p) HCL, $(N, \Theta) = (400, 2.3)$.
(q) HCL, $(N, \Theta) = (100, 2.5)$. (r) HCL, $(N, \Theta) = (200, 2.5)$.
(s) HCL, $(N, \Theta) = (300, 2.5)$. (t) HCL, $(N, \Theta) = (400, 2.5)$.
(u) Algebra. (v) Restaurant-Rev. (w) NDC-classes.
We set $\lambda = 0.5$.}
    \label{box}
\end{figure}
\clearpage
\newpage
\noindent HDMPN achieves an accuracy exceeding 0.2 on all three hypergraph models.
Even at $N = 400$, HDMPN achieves an accuracy above 10\% across different hypergraph models. While HDMP performs similarly well, HDMPN performs notably better than HDMP for the ERH model and slightly but consistently so for the SFH and HCL models. The results for the three empirical hypergraphs, shown in Figure~\ref{accuracy}(f), are consistent with these results for the synthetic hypergraphs.

Figure~\ref{box} shows the performance of the different algorithms in terms of the ranking metric for the synthetic hypergraphs (Figures~\ref{box}(a)--(t)) and empirical hypergraphs (Figures~\ref{box}(u)--(w)). The box plots in the figure show the distribution of the $R$ values of $\mu(s)$ obtained from the $R$ runs. It should be noted that a smaller value of $\mu(s)$ is better. Figure~\ref{box} indicates that HDMPN and HDMP consistently outperform the other methods on all synthetic hypergraphs and most empirical datasets, with HDMPN yielding the best results, followed closely by HDMP. A notable exception arises in the NDC-classes hypergraph, for which HBC is slightly superior to HDMPN and HDMP.

\section{Conclusion and future work}\label{conclusion}

We have proposed the HDMPN algorithm for source detection in a stochastic SI spreading process on hypergraphs. HDMP is a likelihood maximization method and estimates the probability that each node is the source of the spreading process given the final states of all the nodes. With the aim of enhancing predictive accuracy, we have modulated the original likelihood maximization (i.e., HDMP) by a multiplicative factor, which is the fraction of infected neighbors, proposing HDMPN. We have numerically shown that the HDMPN and HDMP outperform three other baseline algorithms and that HDMPN slightly but steadily outperforms HDMP. The latter result further highlights the effectiveness of incorporating the neighbor infection probability.

Despite the promising performance of HDMPN, several limitations must be acknowledged. First, the current framework assumes complete access to the final state of all nodes and homogeneous transmission probabilities across all hyperedges. These assumptions may not hold in realistic situations, where data are often only partially observable or noisy~\citep{zhu2014robust,cheng2024gin}. Second, while it is both analytically and numerically tractable, the present SI model is simplistic and ignores even basic epidemiological factors such as recovery, reinfection, and latent periods. Extending HDMP and HDMPN to accommodate more realistic models, i.e., such as susceptible-infectious-recovered (SIR), susceptible-exposed-infectious-recovered (SEIR), or other hypergraph-specific dynamics, would broaden their applicability in complex spreading scenarios. In particular, the hypergraph structure allows for richer paradigms for modeling epidemic processes. For instance, the critical mass threshold model~\citep{landry2020effect} and the simplicial contagion model~\citep{kiss2023insights} offer alternative mechanisms for capturing group-based interactions and non-linear diffusion effects. It looks possible to extend HDMP and HDMPN to such spreading models on hypergraphs, 
potentially enabling a unified framework adaptable to a variety of propagation mechanisms. Third, we focused on detecting single sources. However, many real-world diffusion processes including epidemic outbreaks may originate from multiple concurrent sources~\citep{jiang2015k,wang2021localization}. Extending HDMP and HDMPN to handle multi-source detection remains an open problem. Finally, both algorithms require $O(N^2 T)$ time. Therefore, developing more scalable algorithms warrants future work. Collectively, these directions among others seed future research.

\backmatter

\bmhead{Acknowledgments}

C.L. is supported by the National Natural Science Foundation of China under Grant No.\ 62473123.
N.M. is supported in part by the NSF under Grant No.\,DMS-2204936, in part by JSPS KAKENHI under Grants No.\,JP21H04595, No.\,23H03414, No.\,24K14840, and No.\,24K030130, and in part by Japan Science and Technology Agency (JST) under Grant No.\,JPMJMS2021.
X.Z. is supported by the China Postdoctoral Science Foundation under Grant No.\,2024M762809.

\bmhead{Data availability}
Data will be made available on reasonable request.

\bibliography{sn-bibliography}

\end{document}